\documentclass[twocolumn,aps,prb,showpacs,amsmath,amssymb]{revtex4-2}
\usepackage{graphicx}
\usepackage{dcolumn}
\usepackage{bm}
\usepackage{xcolor}

\begin{document}

\bibliographystyle{apsrev4-2}

\title{Optimal coupling of HoW$_{10}$ molecular magnets to superconducting circuits near spin clock transitions}

\author{Ignacio Gimeno$^{1,2}$}
\author{V\'{\i}ctor Rollano$^{1,2,3,4}$}
\author{David Zueco$^{1,2}$}
\author{Yan Duan$^{5}$}
\author{Marina C. de Ory$^{6}$}
\author{Alicia Gomez$^{6}$}
\author{Alejandro Gaita-Ari\~{n}o$^{5}$}
\author{Carlos S\'anchez-Azqueta$^{7}$}
\author{Thomas Astner$^{8}$}
\author{Daniel Granados$^{9}$}
\author{Stephen Hill$^{10}$}
\author{Johannes Majer$^{3,4}$}
\author{Eugenio Coronado$^{5}$}
\author{Fernando Luis$^{1,2}$}\email{fluis@unizar.es}

\affiliation{$^{1}$Instituto de Nanociencia y Materiales de Arag\'on, 
CSIC-University of Zaragoza, 50009 Zaragoza, Spain}
\affiliation{$^{2}$Departmento de F\'{\i}sica de la Materia Condensada, 
Universidad de Zaragoza, 50009 Zaragoza, Spain}
\affiliation{$^{3}$Hefei National Laboratory for Physical Sciences at the 
Microscale, University of Science and Technology of China, Hefei 230026, 
China}
\affiliation{$^{4}$Shanghai Branch, CAS Center for Excellence in Quantum 
Information and Quantum Physics, University of Science and Technology of 
China, Shanghai 201315, China}
\affiliation{$^{5}$Instituto de Ciencia Molecular (ICMol), Universidad de 
Valencia, Catedr\'atico Jos\'e Beltr\'an 2, 46980 Paterna, Spain}
\affiliation{$^{6}$Centro de Astrobiolog\'{\i}a (CSIC-INTA), Torrej\'on de 
Ardoz, 28850 Madrid, Spain}
\affiliation{$^{7}$Departamento de F\'{\i}sica Aplicada, Universidad de 
Zaragoza, 50009 Zaragoza, Spain}
\affiliation{$^{8}$Vienna Center for Quantum Science and Technology, 
Atominstitut, TU Wien, 1020 Vienna, Austria}
\affiliation{$^{9}$IMDEA Nanociencia, Cantoblanco, 28049 Madrid, Spain}
\affiliation{$^{10}$National High Magnetic Field Laboratory and Department 
of Physics, Florida State University, Tallahassee, Florida 32310, USA}

\begin{abstract}
A central goal in quantum technologies is to maximize $G$T$_{2}$, where $G$ stands for the coupling of a qubit to control and readout signals and T$_{2}$ is the qubit's coherence time. This is challenging, as increasing $G$ 
(e.g. by coupling the qubit more strongly to external stimuli) often leads to deleterious effects on T$_{2}$. Here, we study the coupling of pure and 
magnetically diluted crystals of HoW$_{10}$ magnetic clusters to microwave 
superconducting coplanar waveguides. Absorption lines give a broadband 
picture of the magnetic energy level scheme and, in particular, confirm the 
existence of level anticrossings at equidistant magnetic fields determined 
by the combination of crystal field and hyperfine interactions. Such 'spin 
clock transitions' are known to shield the electronic spins against magnetic 
field fluctuations. The analysis of the microwave transmission shows that 
the spin-photon coupling becomes also maximum at these transitions. The 
results show that engineering spin-clock states of molecular systems offers 
a promising strategy to combine sizeable spin-photon interactions with a 
sufficient isolation from unwanted magnetic noise sources. 
\end{abstract}


\flushbottom
\maketitle
%
%
\thispagestyle{empty}

\section{Introduction}

Spins embedded in solid hosts are one of the simplest and most natural 
choices to realize qubits, the building blocks of quantum technologies.\cite{Bertaina2007,Awschalom2013} Their quantized spin projections can 
encode the qubit states whereas operations between them can be induced via 
the application of microwave radiation pulses, using well-established 
magnetic resonance protocols. Among the different candidates, chemically 
designed magnetic molecules stand out for several reasons.\cite{Gaita2019,Carretta2021} Besides being microscopic, thus reproducible 
and intrinsically quantum, they represent the smallest structure that remains tuneable. The ability to modify the relevant properties by 
adequately choosing the molecular composition and structure allows 
engineering the qubit spin states and energies.\cite{Martinez-Perez2012,Shiddiq2016} Even more, it enables scaling up computational 
resources within each molecule, e.g. by accommodating several different 
magnetic atoms in exquisitely defined coordinations \cite{Luis2011,Aromi2012,Aguila2014,Ferrando-Soria2016,Fernandez2016} or 
by making use of multiple internal spin states.\cite{Jenkins2017,Godfrin2017,Moreno-Pineda2018,Hussain2018}

This approach however faces the challenge of how to actually implement 
operations and read 
out the results in a realistic device and, even more, how to 'wire up' 
different molecules 
into a scalable architecture. A promising technology is to exploit 
microwave photons 
in circuits, e.g. transmission lines for the control of spin operations 
and resonators for reading out the spin states and for introducing 
effective interactions.\cite{Majer2007,Schoelkopf2008,Jenkins2013,Jenkins2016,Bonizzoni2018,Rollano2022} 
Working with high-spin molecules helps maximizing the spin-photon 
coupling, as required for such applications.\cite{Jenkins2013} However, it also tends to enhance decoherence, as their interactions with fluctuating 
hyperfine and dipolar magnetic fields also become stronger.\cite{Morello2006,Escalera-Moreno2019} 

\begin{figure}
\centering
\includegraphics[width=\columnwidth]{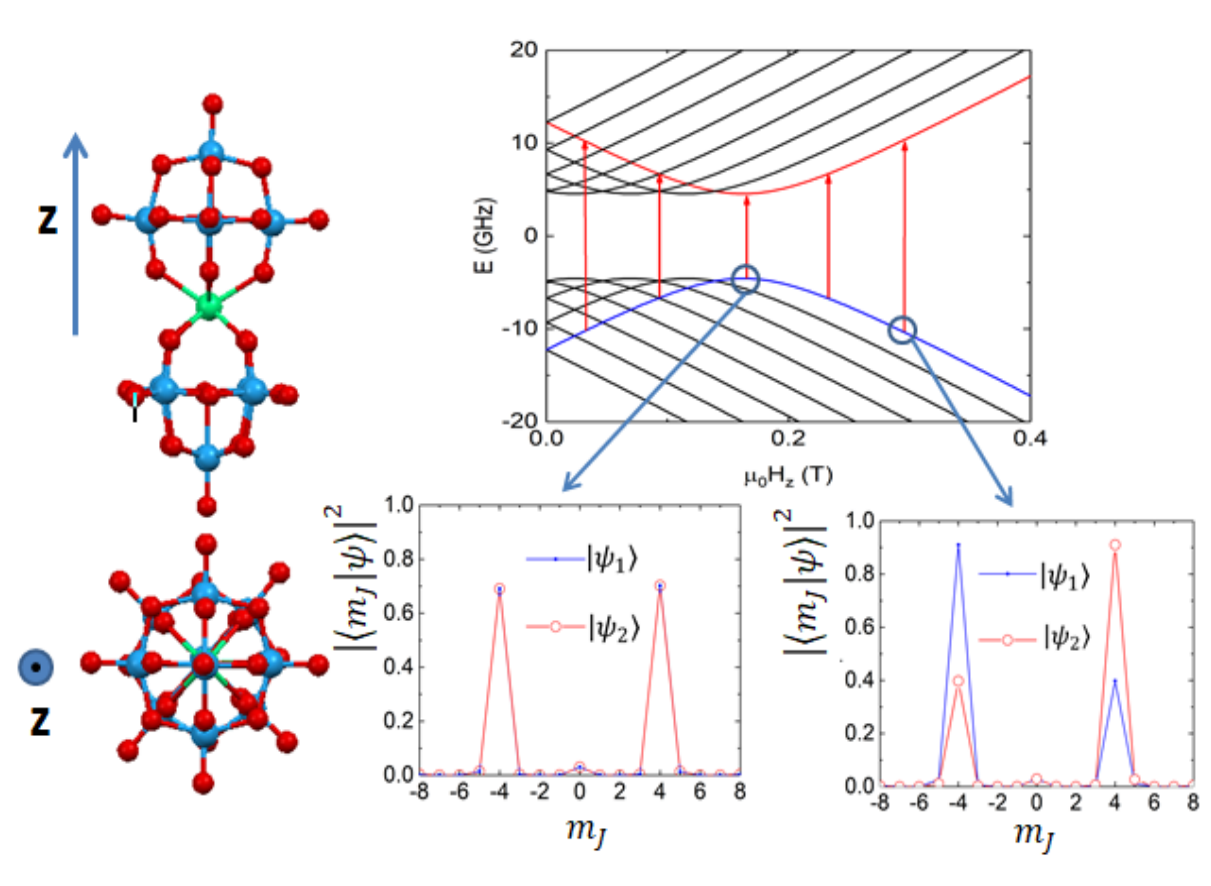}
\caption{Left: two views of the HoW$_{10}$ cluster, showing its 
fourfold coordination symmetry around its anisotropy axis $z$. Right: scheme of magnetic energy levels corresponding 
to the electronic ground doublet ($m_J=\pm 4$) of HoW$_{10}$ and wave functions of two mutually 
avoiding levels calculated at the indicated magnetic fields (both at and slightly off the clock 
transition).}
\label{Levels}
\end{figure}

A general strategy to reconcile a high qubit density 
with sufficient isolation is to encode each qubit in 
states that are robust against the dominant noise sources.\cite{Lidar1998} This idea 
underlies the  design of the trasmon superconducting qubit\cite{Koch2007} and of 
several semiconducting quantum dot qubits.\cite{Burkard2022} In the case of spins, isolation from magnetic field 
fluctuations can be achieved by associating $0$ and $1$ to superpositions states that 
arise at avoided level crossings, or 'spin-clock' transitions.\cite{Longdell2006,McAuslan2012,Wolfowicz2013}
Such transitions have been observed in impurity dopants in semiconductors\cite{Wolfowicz2013} and in crystals hosting lanthanide 
ions,\cite{Longdell2006,McAuslan2012} and can arise from either significant off-diagonal anisotropy 
terms in non-Kramers electronic spins or from hyperfine couplings in 
electronuclear spin systems. They have also recently been studied in 
magnetic molecules.\cite{Shiddiq2016,Zadrozny2017,Collett2019,Rubin-Osanz2021,Gimeno2021,Kundu2022} A paradigmatic example of the latter is provided by the sodium 
salt of the cluster [Ho(W$_{5}$O$_{18}$)$_{2}$]$^{9-}$,\cite{AlDamen2009,Ghosh2012} 
hereafter referred to as HoW$_{10}$, which consists of a single 
Ho$^{3+}$ ion encapsulated by polyoxometalate moieties (Fig. \ref{Levels}). Its fourfold coordination 
symmetry gives rise to 
fourth order off-diagonal terms in the spin Hamiltonian that 
strongly mix the $m_{J} = \pm 4$ projections of the ground 
electronic spin doublet. The large quantum tunneling gap $\Delta \simeq 9.1$ GHz generated by such terms, combined with the hyperfine 
interaction with the $I=7/2$ spin of the Ho nucleus, gives rise to a 
set of level anti-crossings (see Fig. \ref{Levels}). Near each of them, the spin coherence time T$_{2}$ is sharply enhanced \cite{Shiddiq2016} and the electron spin system effectively decouples from the nuclear spin. \cite{Kundu2023}

\begin{figure}
\centering
\includegraphics[width=\columnwidth]{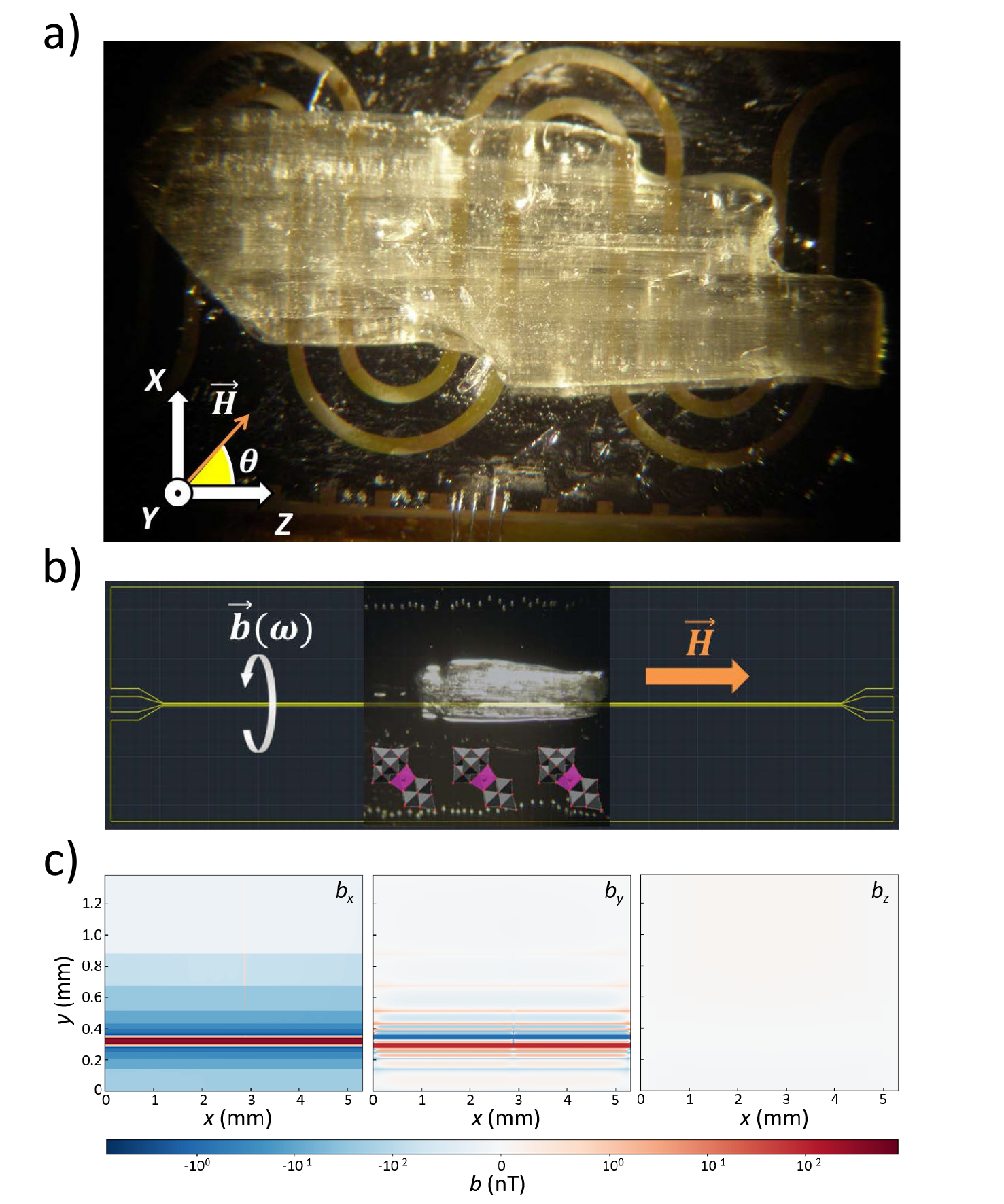}
\caption{a) Image of a chip with a meander-like $400$ $\mu$m wide superconducting 
transmission line hosting a single crystal of pure HoW$_{10}$. This device was 
used for angle-dependent experiments performed at $T=4.2$ K, with $X$, $Y$ and $Z$ being the axes of 
the laboratory reference frame. b) Sketch of a chip with a
$35$ $\mu$m wide straight superconducting line hosting a single crystal of Ho$_{0.2}$Y$_{0.8}$W$_{10}$. 
The inset shows the approximate orientation of the 
molecules in the crystal. This device was used in very low-$T$ experiments. c) $2D$ plots of the 
microwave field $\vec{b}$ components along the three laboratory 
axes in the area defined by the crystal shown in (b). This magnetic field 
generates transitions between different spin states. It was calculated 
numerically by means of finite-element methods. The results show that $\vec{b}$ 
is confined in a plane perpendicular to the line and in a region very close to 
it.}
\label{chip}
\end{figure}

In this work, we explore the coupling of Ho$_{x}$Y$_{1-x}$W$_{10}$ 
single crystals ($x=0.2$ and $1$) to superconducting co-planar 
waveguides. These experiments provide a direct method to 
investigate in detail how the spin-photon coupling evolves as a 
function of magnetic field, thus both near and far from the spin-clock transitions, and temperature. The manuscript is organized as follows. Section II provides details 
on the preparation of 
the samples, the design and fabrication of the devices and the 
transmission measurements. 
Section III describes results obtained under different 
experimental conditions and discusses them with the help of input-output theory. The last 
section IV is left for the conclusions. \\

\section{Experimental details}

\subsection{Sample preparation and characterization}
The synthesis of Ho$_{x}$Y$_{1-x}$W$_{10}$ crystals followed established 
protocols.\cite{AlDamen2009} The samples were kept in their mother 
solution until a experiment had to be performed, in order to protect them 
from degradation. The samples were characterized by means of specific 
heat and magnetic measurements. The results agree with those reported 
previously \cite{AlDamen2009} and therefore confirm that HoW$_{10}$ 
clusters have an $m_{J} = \pm 4$ electronic spin ground 
state and a sizeable quantum tunneling gap $\Delta$.

\begin{figure}
\centering
\includegraphics[width=\columnwidth]{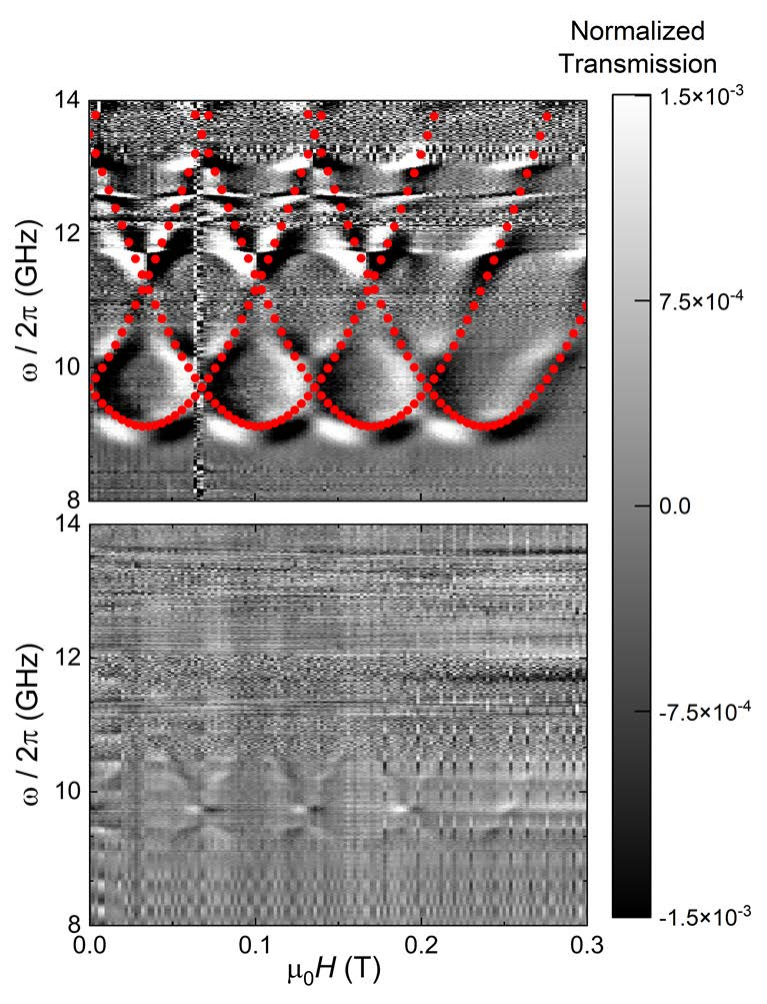}
\caption{Normalized transmission through a $400$ $\mu{\rm m}$ wide 
superconducting line coupled to a single crystal of pure HoW$_{10}$ (top 
panel) and of Ho$_{0.2}$Y$_{0.8}$W$_{10}$ (bottom panel). In both cases, the normalization (Eq. (\ref{normalization})) was done with $\mu_{0} \Delta H = 2$~mT. The experiments 
were performed at $4.2$ K and the magnetic field was parallel to the $Z$ 
laboratory axis (see Fig. \ref{chip}a). The dotted lines in the top panel 
show the frequencies of allowed spin 
transitions, derived from Eq. (\ref{Hamiltonian}).} 
\label{2Dplots_4K2}
\end{figure}

\subsection{Superconducting device design and fabrication}
Two types of circuits hosting superconducting coplanar waveguides were employed in the microwave transmission experiments that form the core of this work. The 
first one consists of a $400 \mu$m wide 
central transmission line separated from two ground planes by $200 \mu$m wide gaps. It was fabricated 
by optical lithography of $100$ nm thick Nb films deposited by sputtering 
onto a single-crystalline sapphire substrate. The size of the central line and 
its meander shape were designed in order to match the 
dimensions (ca. $10 \times 4 \times 1$ mm$^{3}$) of the HoW$_{10}$ single crystals that were measured in experiments performed at $4.2$ K (Fig. \ref{chip}a). 

\begin{figure}
\centering
\includegraphics[width=\columnwidth]{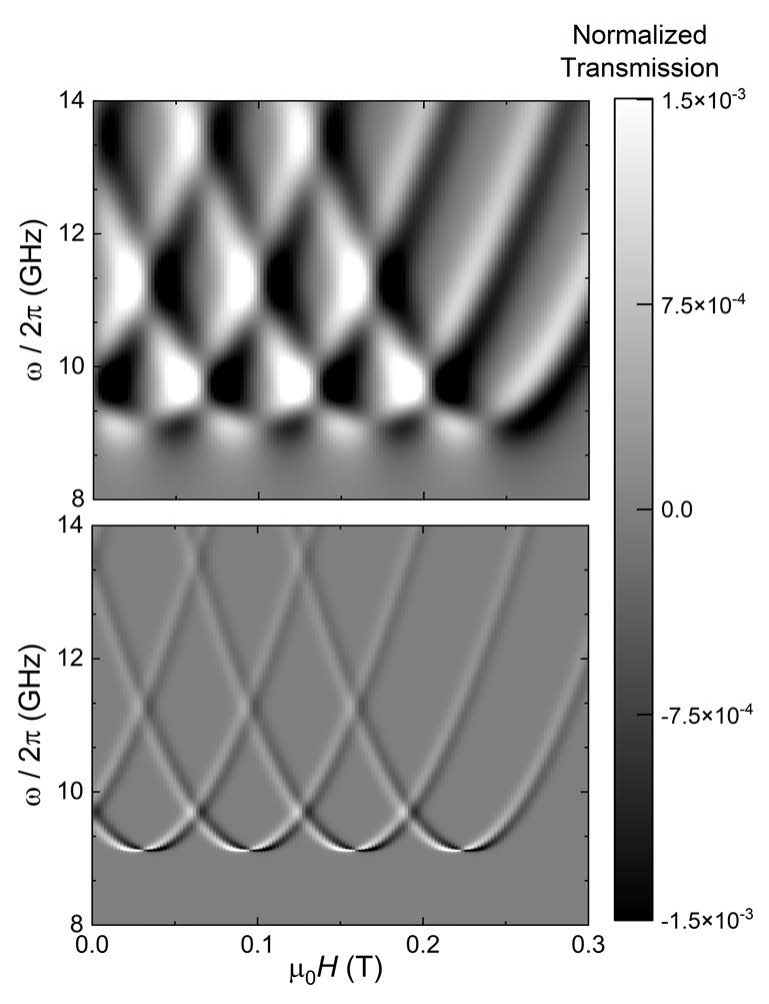}
\caption{Simulated $2D$ plots of the normalized transmission through a $400$ $\mu$m wide 
superconducting line coupled to a single crystal of pure HoW$_{10}$ (top 
panel) and of Ho$_{0.2}$Y$_{0.8}$W$_{10}$ (bottom panel) and for a magnetic field applied along the $Z$ laboratory axis. The 
results are calculated averaging Eq. (\ref{eq:transmission-line}) , with the spin-photon coupling $G$ given by Eq. (\ref{Gamma}) and $g^{2}(\omega_{12}) = \alpha \omega_{12}$, over a gaussian distribution of bias fields with a width $\sigma = 6.4$ mT for $x=1$ and $\sigma = 2.7$ mT for $x=0.2$. The fitting 
parameters were the angle $\simeq 45$ degrees that $\vec{H}$ makes with respect to the magnetic easy axis $z$ and the dimensionless $\alpha \simeq 3.2 \times 10^{-4}$ for HoW$_{10}$ and $\alpha \simeq 8 \times 10^{-6}$ for Ho$_{0.2}$Y$_{0.8}$W$_{10}$.} 
\label{2Dplots_4K2_theo}
\end{figure}

A second device was fabricated to optimize the coupling to the smaller size,  
magnetically diluted Ho$_{x}$Y$_{1-x}$W$_{10}$ crystals employed in the very low-T experiments (Fig. \ref{chip}b). It consists of a $35 \mu$m wide straight transmission line, separated from the 
ground planes by $20 \mu$m wide gaps in order to maintain a $50$ Ohm characteristic 
impedance. It was fabricated by maskless lithography and reactive ion etching 
techniques on a $100$nm thick Nb film deposited by means of DC magnetron sputtering on a $275 \mu$m thick silicon substrate. The native oxide of the 
silicon wafer was previously removed using a hydrofluoric acid bath. The base pressure prior to the deposition of Nb was better than $2 \times 10^{-8}$ Torr.

The magnetic field distribution generated by the microwave superconducting currents propagating via these transmission lines has been calculated using the 
electromagnetic simulation package SONNET \cite{SONNET} and finite element simulations. Results of these simulations are shown in Fig. \ref{chip}c. The microwave  
magnetic field is confined in a plane perpendicular to the line. This information is relevant to prepare and interpret the transmission experiments, because resonant 
transitions between different spin states are only allowed if the microwave field has a 
non zero projection along the molecular magnetic anisotropy axis $z$ (see section III for 
details). Besides, its magnitude falls off quickly as one moves away from the line. This means that the experiments 
typically explore the coupling of a very small region of the crystal. Working with small crystals and  
sufficiently small lines, which becomes feasible at very low temperatures, helps to mitigate the effects of inhomogeneities associated with crystal 
twinning.         

\subsection{Microwave transmission experiments}
The crystals were attached onto the transmission line with apiezon N grease. Microwave transmission experiments were performed by connecting 
the input and output ports of the chip to a vector network analyzer that measures the transmission coefficient $S_{21}$ for frequencies $\omega/2 \pi$ ranging between $0.01$~GHz and $14$~GHz. For experiments at $T = 4.2$ K, 
the chips were submerged in the liquid Helium bath of a cryostat 
equipped with a $9$ T $\times 1$ T $\times 1$ T vector magnet. This 
set-up allows applying dc magnetic fields with amplitudes 
$\mu_{0}H$ up to 
$1$~T along any arbitrary direction in the $X$, $Y$, $Z$ laboratory 
reference frame shown in Fig. \ref{chip}a. In these experiments, 
$\vec{H}$ was rotated within the 
$X-Z$ plane of the device. Experiments were also performed at 
temperatures below $1$ K, from $50$ mK up to $800$ mK, in order to 
control and optimize the thermal polarization difference $\Delta P_{12}$ 
between the levels involved in each resonant transition, thus the spin-photon 
coupling. The chips were thermally anchored to the 
mixing chamber of a cryo-free dilution refrigerator, and placed at the 
centre of an axial $1$ T magnet ($\vec{H}$ was parallel to $Z$ in this 
case, as shown in Fig. \ref{chip}b). The transmission experiments were performed as described above, 
with the inclusion of a set of attenuators, for a total $-50$ dB, in 
the input line and of a low noise cryogenic amplifier (gain $\simeq +35$
dB) at the $T = 4$ K stage in the output line. 

In order to compensate for the decay of the waveguide transmission with 
increasing frequency and to enhance the contrast of those effects 
associated with its coupling to the spins, $S_{21}$ was 
normalized. For this, we compare transmission data measured at two different magnetic fields. \cite{Clauss2013} The normalized transmission $t$ at 
magnetic field $H$ and frequency $\omega$ is given by 

\begin{equation}
t(H,\omega) = \frac{S_{21}(H,\omega)-S_{21}(H+\Delta H,\omega)}{S_{21}^{(0)}(\omega)}
\label{normalization}
\end{equation}

\noindent where $\Delta H > 0$ and $S_{21}^{(0)}$ is the transmission 
of the 'bare' transmission line. In practice, $S_{21}^{(0)}$ is 
measured at a magnetic field for which all spin excitations lie outside 
the accessible frequency region. For $\Delta H$ smaller than the magnetic 
field width of a given spin transition, 
$t$ approximately corresponds to the derivative of the normalized 
transmission, similar to the signal detected in conventional Electron 
Paramagnetic Resonance (EPR) experiments. The actual transmission can 
also be obtained, by choosing a larger $\Delta H$ 
but at the cost of deteriorating the signal-to-noise ratio.

\begin{figure*}
\centering
\includegraphics[width=\textwidth]{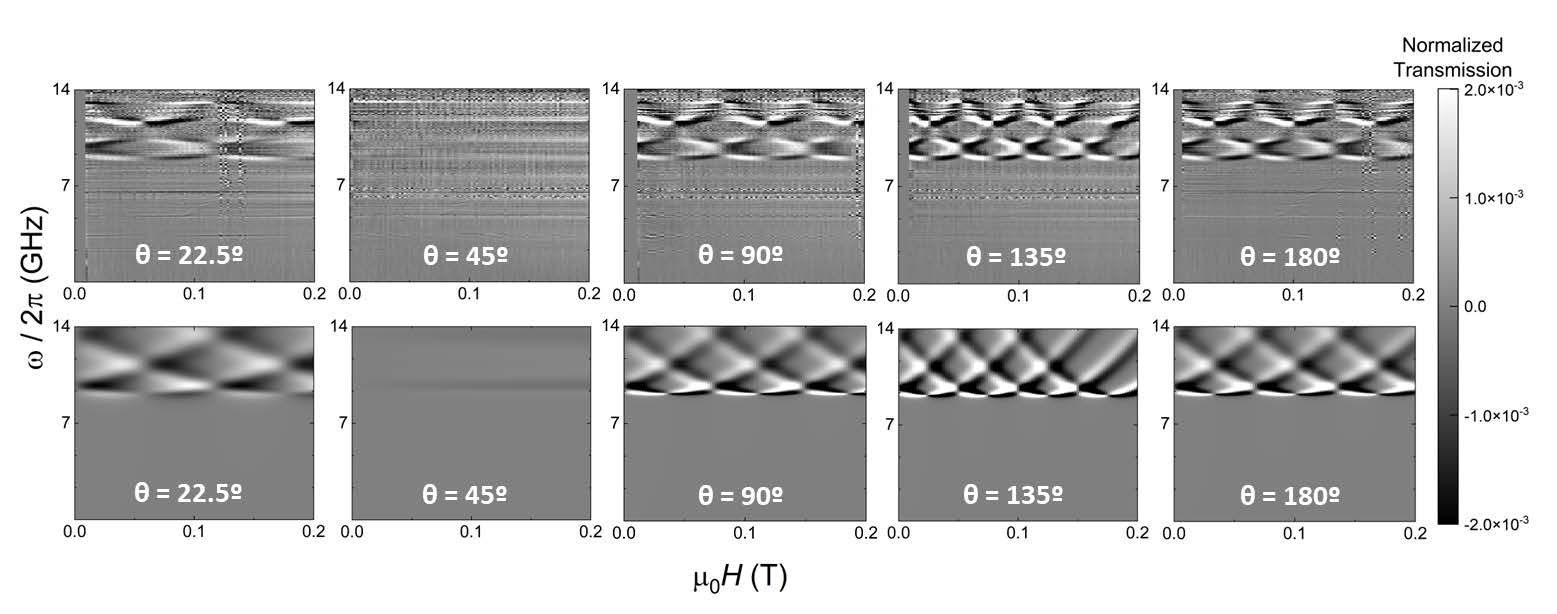}
\caption{Experimental (top) and simulated (bottom) normalized transmission through a $400$ $\mu$m wide 
superconducting line coupled to a single crystal of HoW$_{10}$ for 
different orientations of the magnetic field. The experiments were 
performed at $T=4.2$ K. The normalization of the experimental data (Eq. (\ref{normalization})) was done with $\mu_{0} \Delta H = 2$ mT. The simulations are calculated averaging Eq. (\ref{eq:transmission-line}), with the spin-photon coupling $G$ given by Eq. (\ref{Gamma}) and $g^{2}(\omega_{12}) = 3.2 \times 10^{-4} \omega_{12}$, over a $\sigma = 6.4$ mT wide gaussian distribution of bias fields.}
\label{2Dplots_angle}
\end{figure*}

\section{Results}

\subsection{Broad-band spectroscopy: field-tuned clock transitions}

Figure~\ref{2Dplots_4K2} shows two-dimensional maps of the transmission 
through a $400$ $\mu$m wide transmission line coupled to large pure 
(top panel) and magnetically dilute (bottom panel) crystals. These data 
were measured at $T = 4.2$ K with the dc magnetic field applied along the 
$Z$ axis. Because of the geometry of the line (see image in Fig. \ref{chip}a and simulations in \ref{chip}c), 
the microwave magnetic field felt by the crystal was mainly confined to 
the $Y-Z$ plane. The data neatly show changes in transmission associated 
with the resonant absorption of microwave photons by the HoW$_{10}$ spins. Each of 
these resonances corresponds to an allowed transition between two states 
with a different electronic spin state and the same nuclear spin state, 
such as those marked by vertical arrows in 
Fig. \ref{Levels}. These resonance lines provide then a complete 
picture of the low-lying magnetic energy levels in HoW$_{10}$. In 
particular, they show the presence of a finite gap $\Delta \simeq 9.1$ 
GHz in the excitation spectrum at $4$ different avoided level crossings. 
The spectroscopic patterns of pure and magnetically diluted crystals 
agree, save for the narrower lines observed in the latter case and the difference in absorption intensities associated with 
the number of spins that effectively couple to the propagating 
photons in each case. 

\subsection{Numerical simulation of the transmission spectra}

The $\omega(H)$ dependence of the different resonance lines can be 
estimated from the spin Hamiltonian \cite{AlDamen2009,Ghosh2012}

\begin{equation} 
\mathcal{H}=B_{20}\hat{O}_2^0+B_{40}\hat{O}_4^0+B_{60}\hat{O}_6^0+B_{44}\hat{O}_4^4 + g_{J} \mu_{\rm B}\vec{H} \cdot \vec{J} + AJ_{z}I_{z}
\label{Hamiltonian}
\end{equation} 

\noindent that includes four crystal field terms, the Zeeman interaction 
with the external magnetic field and the hyperfine interaction. The 
parameters $g_{J}=5/4$, $B_{20}=0.601$ cm$^{-1}$, $B_{40}=6.93 \times 10^{-3}$ cm$^{-1}$, $B_{60}=-5.1 \times 10^{-5}$ cm$^{-1}$, $B_{44}=3.14 \times 10^{-3}$ cm$^{-1}$,  $A=2.77 \times 10^{-2}$ cm$^{-1}$ have been 
determined from EPR experiments on magnetically diluted 
samples.\cite{Ghosh2012,Shiddiq2016} It follows that the ground state 
corresponds to the $m_{J} = \pm 4$ doublet, split by tunneling terms (mainly 
the $B_{44} \hat{O}_{4}^{4}$ term) and by hyperfine interactions. Because of 
the very strong uniaxial magnetic 
anisotropy of HoW$_{10}$, avoided level crossings occur at $H_{z} \simeq 2 m_{I} H_{z,1}$, with $H_{z,1} = 23$ mT for the crossing of states with nuclear 
spin projection $m_{I}=1/2, 3/2, 5/2$ and $7/2$, respectively. The only 
free parameter is then the orientation of the molecular easy axis $z$ 
with respect to the external magnetic field, which amounts to rescaling 
the magnetic field axis. As shown in Fig. \ref{2Dplots_4K2}, we find a 
good agreement with the same parameters given above. 
These results show that the concentrated crystals used in this work retain the same magnetic 
anisotropy and confirm that the strong spin tunneling, and the associated 
energy gap, are genuine properties of each molecule. 

It is also possible to simulate the full transmission spectra. For this, 
we apply input-output theory to the interaction of microwave photons 
propagating via the transmission line with the electronic magnetic 
moments of the molecules. The complex transmission is then given by\cite{Fan2010,Burillo2016}  

\begin{equation} 
S_{21}^{*} = 1- \frac{G} {G + \gamma + i (\omega_{12}-\omega)} 
\label{eq:transmission-line}
\end{equation}
   
\noindent where $G$ is the photon-induced transition rate between spin states $\vert \psi_{1} \rangle$, with energy $E_{1}$, and $\vert \psi_{2} \rangle$, with energy $E_{2}$, (see 
Fig. \ref{Levels}), $\gamma$ is the spin line width and $\omega_{12} = (E_{2}-E_{1})/\hbar$ is the resonance frequency at the given magnetic 
field. The interaction constant $G$ parameterizes the spin photon coupling and is, therefore, our main interest in this work. Time dependent perturbation theory gives the following expression

\begin{equation}
G \simeq 2 \pi g^{2}(\omega_{12}) \vert \langle \psi_{1} \vert J_{z} \vert \psi_{2} \rangle \vert^{2} \left[ n(\omega_{1,2})+1 \right] \Delta P_{12}
\label{Gamma}
\end{equation}

\noindent where $g(\omega_{12})$ is a spin-photon coupling density, which depends on the mode density in the transmission line and on geometrical 
factors (mainly the number of spins, their location with respect to the circuit and 
the latter's geometry), $n(\omega_{1,2}) = \left[ \exp {\left( h \omega_{1,2} / k_{\rm B} T \right)} - 1 \right]^{-1}$ is the bosonic 
occupation number, $\Delta P_{12} = \left[\exp{(-E_{1}/k_{\rm B}T)}-\exp{(-E_{2}/k_{\rm B}T)}\right]/Z$ is the thermal population difference between the two levels and $Z$ is the partition function. Only the microwave field component parallel to the 
anisotropy axis $z$ contributes to the coupling, because 
$\langle \psi_{1} \vert J_{x,y} \vert \psi_{2} \rangle = 0$ for any 
superposition of $\vert m_{J} = \pm 4 \rangle$ states. This 
explains why $G$ is determined by the matrix element of $J_{z}$.

We have performed numerical simulations of the normalized transmission 
amplitude based on Eqs. (\ref{normalization}), (\ref{eq:transmission-line}) and (\ref{Gamma}) 
and the spin wave functions derived from the spin Hamiltonian (\ref{Hamiltonian}). We 
approximate the spin-photon coupling density by the expression $g^{2}(\omega_{12}) \simeq \alpha \omega_{12}$, valid in the limit of one-dimensional transmission lines.\cite{GarciaRipoll2022} Here, $\alpha$
is an adjustable fitting parameter. The resonance line widths are of order  $100-300$ MHz for both the pure and magnetically diluted crystals. Electron 
spin resonance experiments performed on Ho$_{0.2}$Y$_{0.8}$W$_{10}$ show that T$_{2} \sim 0.1$ $\mu$s at $T = 4.2$ K. The homogeneous broadening $\sim 10$ MHz is 
therefore much smaller than the resonance width observed in experiments, suggesting that the latter is dominated by the 
inhomogeneous broadening. We recognize that there are 
multiple sources of line broadening,\cite{Shiddiq2016} which include dipole-dipole interactions between 
molecular spins and distributions in the orientations of the molecular axes 
and of their crystal field parameters (e.g. $B_{4}^{4}$ that gives rise to 
the tunnel splitting $\Delta$).\cite{Ghosh2012, Shiddiq2016} In the 
simulations shown in Fig. \ref{2Dplots_4K2_theo} their effect was introduced 
with gaussian field distributions having $\sigma \simeq 6.4$ mT for $x=1$ and $\sigma \simeq 2.7$ mT for $x=0.2$. These parameters were chosen to give 
the best overall agreement with the experiments. 

\begin{figure}
\centering
\includegraphics[width=\columnwidth]{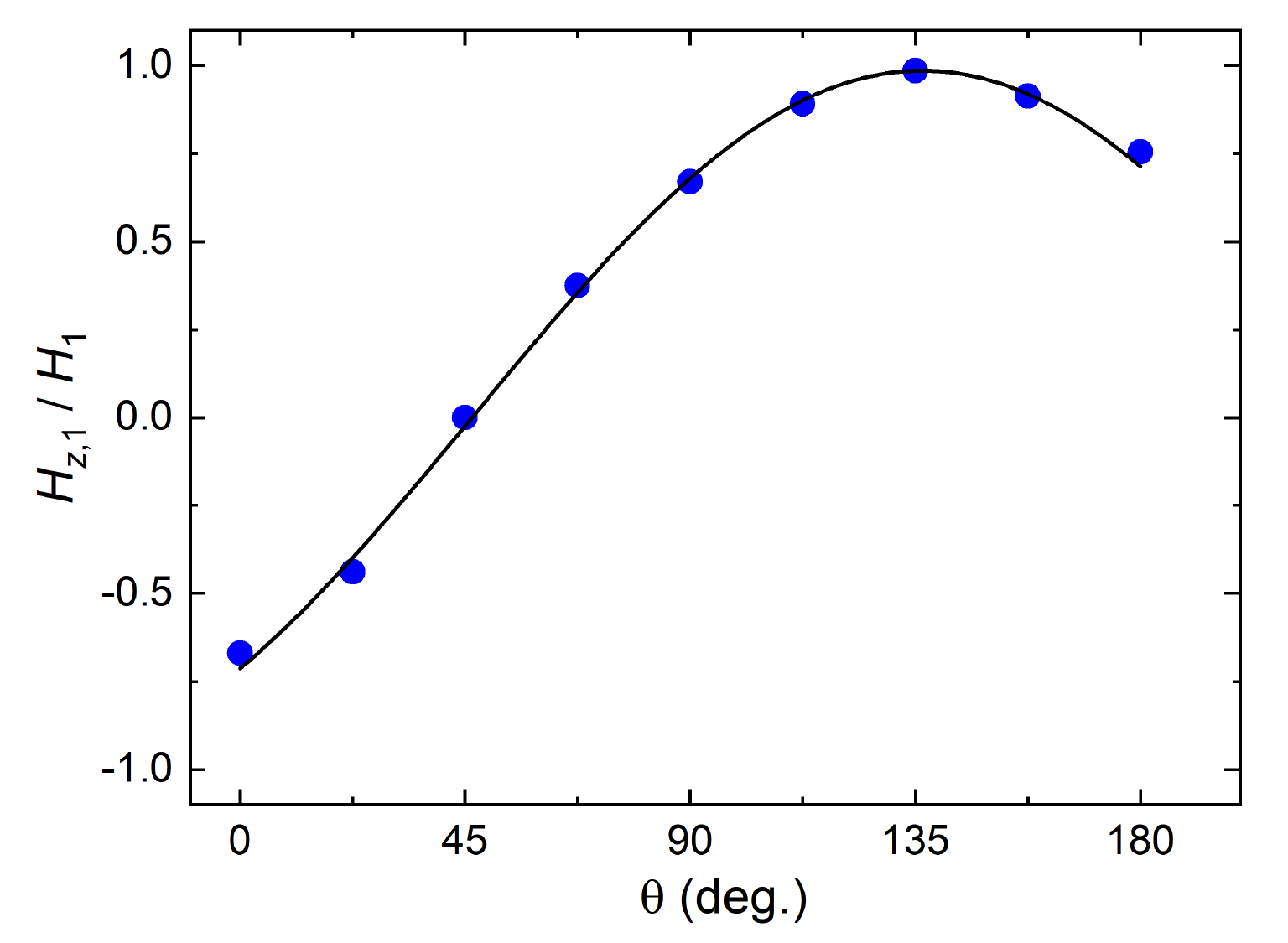}
\caption{Magnetic field projection along the 
molecular magnetic anisotropy axis $z$ as a function of the 
orientation of 
$\vec{H}$ within the laboratory $XZ$ plane (see Fig. \ref{chip}a). 
Here, $H_{1}$ is the magnetic field at which the first avoided 
level crossing is experimentally observed (Fig. \ref{2Dplots_angle}) and $H_{z,1} = 23$ mT is the first crossing 
longitudinal field derived from the spin Hamiltonian (\ref{Hamiltonian}).}
\label{Hz1vsangle}
\end{figure}

\subsection{Dependence on magnetic field orientation}

The experiments on the pure HoW$_{10}$ crystal (Fig. \ref{chip}a) were repeated for different orientations of $\vec{H}$ in the $XZ$ plane of the 
chip. This geometry allows varying the angle between $\vec{H}$ and 
the molecular anisotropy axis $z$, while minimizing effects associated with the excitation and motion of 
superconducting vortices. The results are shown in Fig. \ref{2Dplots_angle}. 
Clear changes in the absorption pattern are observed. They correspond to different magnetic field periodicities of the avoided level crossings. As we have mentioned above, these anticrossings are mainly determined by the condition $H_{z} = 2m_{I} H_{z,1}$, which requires reaching higher 
magnetic fields strengths the more $\vec{H}$ deviates from the 
anisotropy axis. When the magnetic field forms an angle $\theta = 45$ 
degrees with the $X$ and $Z$ laboratory axes, the pattern 
disappears, showing that $\vec{H}$ is then nearly orthogonal to 
$z$. By contrast, the pattern period $H_{1}$ becomes minimum 
for $\theta \simeq 135$ degrees, showing that $\vec{H}$ is then  closest to $z$ within the $X-Z$ plane.  
The dependence of the experimental $H_{1}$ on $\theta$ is shown in Fig. \ref{Hz1vsangle}. Fitting these data allows estimating in situ the orientation of the 
magnetic anisotropy axis $z$ with respect to the crystal and to the laboratory reference frame. The results are compatible with $z$ pointing along the long molecular axis (Fig. \ref{Levels}). 

\begin{figure}
\centering
\includegraphics[width=\columnwidth]{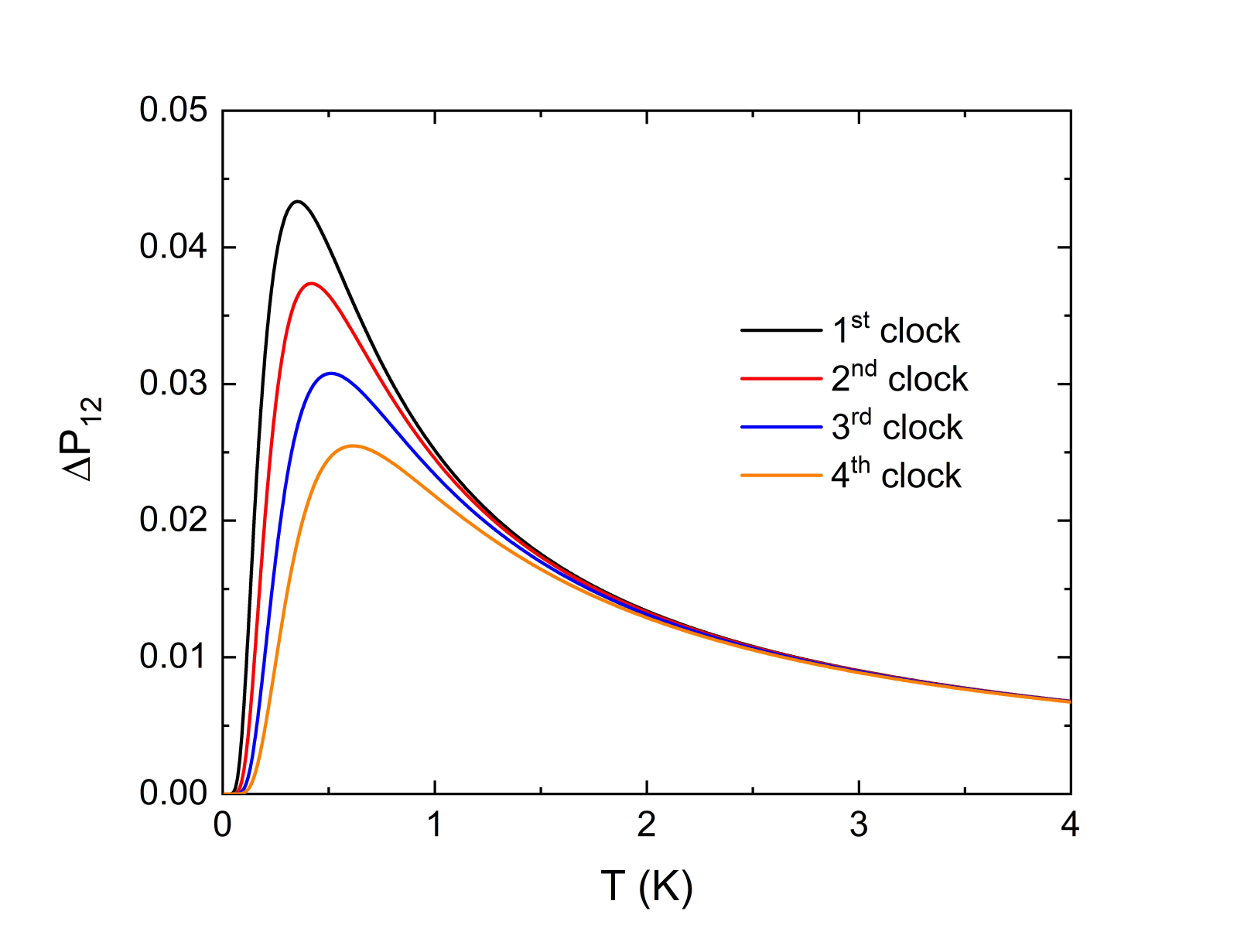}
\caption{Thermal equilibrium population difference between the two levels involved in each of the spin-clock transitions in HoW$_{10}$.}
\label{DeltaPvsT}
\end{figure}

Once the orientation of $z$ is set, the positions of the 
resonances and the full transmission spectra can be calculated for 
any magnetic field angle. The results, shown in Fig. \ref{2Dplots_angle}, agree very well with the experimental ones. 
This agreement confirms the very strong uniaxial magnetic
anisotropy of HoW$_{10}$ and provides a basis to analyze how 
the spin photon coupling depends on temperature and magnetic field 
strength. Besides, it shows that $\vec{b}$ generated by a straight 
transmission line (Fig. \ref{chip}b and \ref{chip}c), although 
perpendicular to the external $\vec{B}$, has a sizeable component 
along $z$. Therefore, it should also provide a nonzero spin-photon coupling. This simpler 
geometry (Fig. \ref{chip}b) was then adopted for experiments 
performed at very low temperatures, which are discussed in what follows.   

\begin{figure*}
\centering
\includegraphics[width=\textwidth]{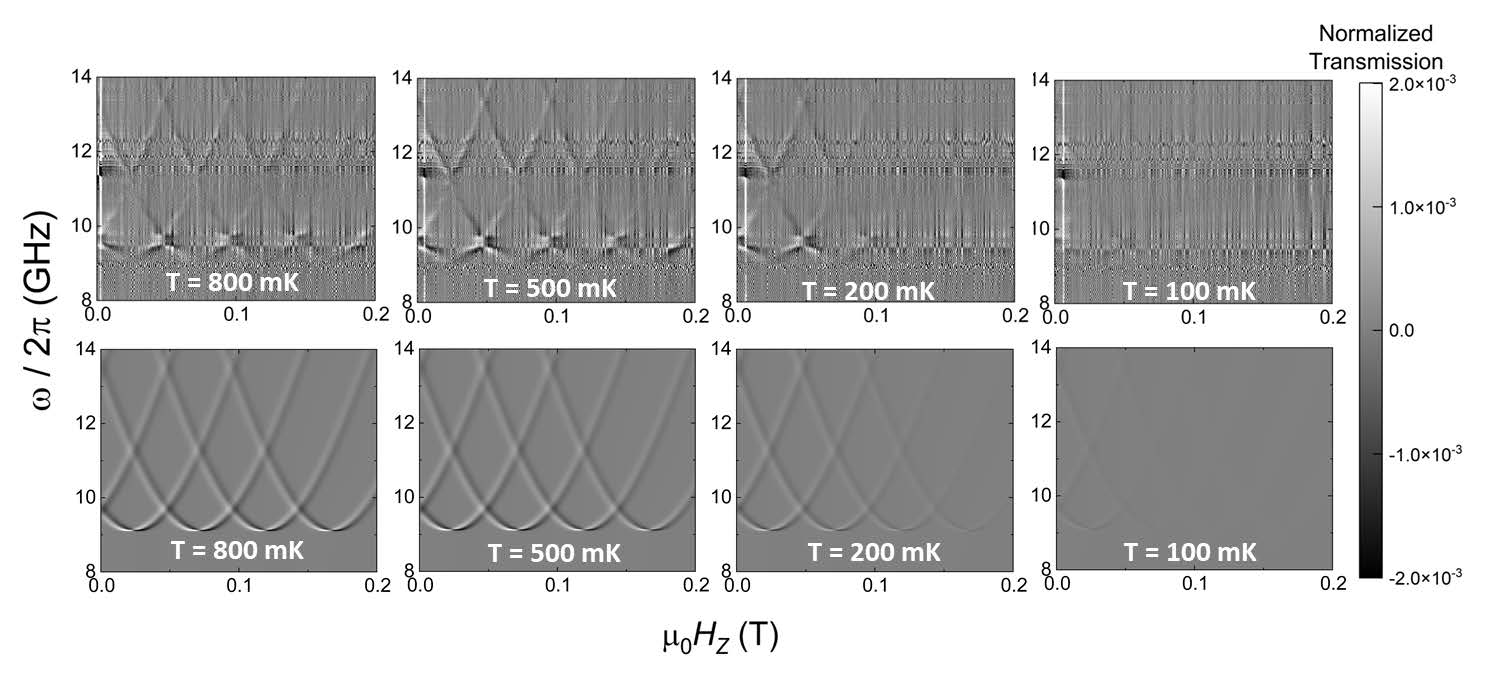}
\caption{Experimental (top) and simulated (bottom) normalized transmission through a $35$ $\mu$m wide 
superconducting line coupled to a single crystal of Ho$_{0.2}$Y$_{0.8}$W$_{10}$ for 
different temperatures. The magnetic field was applied along the $Z$ laboratory axis (Fig. \ref{chip}b). The normalization of the experimental data (Eq. (\ref{normalization})) was done with $\mu_{0} \Delta H = 1$ mT. 
The simulations are calculated averaging Eq. (\ref{eq:transmission-line}), with the spin-photon coupling $G$ given by Eq. (\ref{Gamma}) and $g^{2}(\omega_{12}) = 1.6 \times 10^{-5} \omega_{12}$, over a $\sigma = 2.7$ mT wide gaussian distribution of bias fields.}
\label{2DvsT}
\end{figure*}

\subsection{Broad band spectroscopy below 1 K: temperature dependence of the spin-photon coupling}

The relative populations of the spin levels involved in a 
resonant transition influence the effective spin-photon 
coupling $G$ (see Eq. (\ref{Gamma})). In equilibrium, this  
introduces a temperature dependence through the polarization 
parameter $\Delta P_{12}$, which is plotted in Fig. \ref{DeltaPvsT}. Decreasing $T$ leads to a larger 
polarization provided that $k_{B}T$ remains sufficiently high with respect to $\Delta$ and to the hyperfine splitting of each 
electronic level. The fact that spin-clock 
transitions in HoW$_{10}$ involve two excited levels gives 
rise to a maximum followed by a rapid drop in 
polarization. 

This behaviour is confirmed by experiments performed with the 
circuit shown in Fig. \ref{chip}b on a  
Ho$_{0.2}$Y$_{0.8}$W$_{30}$ single crystal. Transmission spectra measured at different temperatures are shown 
in Fig. \ref{2DvsT}. The relative intensities of the four 
clock transitions remain comparable to each other until, on 
cooling below $0.5$ K, they begin to gradually fade away  
from right ($n = 4$) to left ($n = 1$). Numerical calculations 
based on Eqs. (\ref{eq:transmission-line}) and (\ref{Gamma}) are 
also shown in Fig. \ref{2DvsT}. They agree 
with this behaviour. For this reason, we 
have chosen the data measured at $T = 0.65$ K to study the 
magnetic field dependence of the spin-photon coupling.

\subsection{Magnetic field dependence of the spin-photon coupling near spin-clock transitions}

Whereas the positions of the resonance lines give access to the energy level scheme, 
their intensities provide information on the wavefunctions of the involved states. An 
important advantage of working with open waveguides is that both frequency and 
magnetic field can be varied independently of each other. It is therefore possible to 
monitor how the absorption intensity varies as a function of $H$. 

\begin{figure}
\centering
\includegraphics[width=\columnwidth]{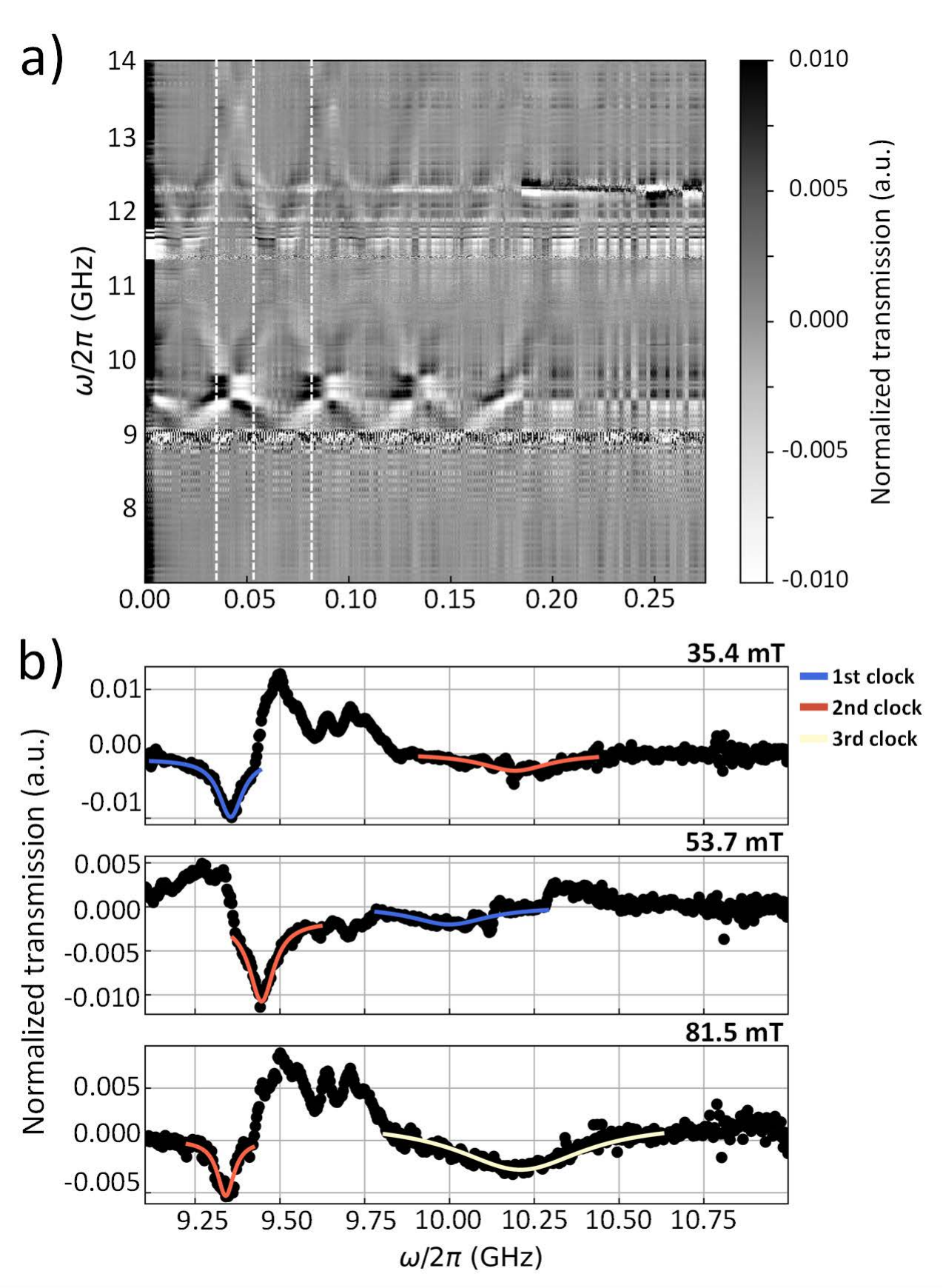}
\caption{a) 2D plot of the transmission measured, at $T=0.65$ K, on a $35$ $\mu$m 
wide transmission line coupled to a single crystal of Ho$_{0.2}$Y$_{0.8}$W$_{10}$. The normalization of the experimental data (Eq. (\ref{normalization})) was done with $\mu_{0} \Delta H = 15$ mT. b) Transmission data as a function of frequency at the 
fields marked by vertical dotted lines in (a). The spin absorption lines correspond to the transmission dips, whereas the maxima correspond to (minus) the absorption lines of the data used in the normalization. Additional 'bumps' are also visible. They arise from spurious transmission modes, which give rise to horizontal lines in the 2D plot of panel a). The solid lines are fits, based on Eq. (\ref{eq:transmission-line}), of the different 
resonances from which the spin-photon coupling $G$ and the line width $\gamma$ are determined.}
\label{2Dplot_650mK}
\end{figure}

Figure \ref{2Dplot_650mK}a shows a $2D$ plot of normalized transmission data 
measured 
at $T = 0.65$ K. As discussed above (see also Fig. \ref{DeltaPvsT}), this 
temperature provides a good polarization $\Delta P_{12}$ for all relevant 
spin transitions. The normalization (Eq. (\ref{normalization})) was performed by subtracting data 
measured at magnetic fields separated by $\mu_{0} \Delta H = 15$ mT. The 
minima in the normalized transmission traces (Fig. \ref{2Dplot_650mK}b) provide then the full absorption 
resonance lines at the given fields, whereas the maxima 
correspond to (minus) the absorption at $H+\Delta H$. Notice that the relative positions of minima and maxima reflect the magnetic field slope of 
the HoW$_{10}$ transition frequencies. Spurious resonant modes of the 
transmission line lead to additional transmission 'bumps' that form 
horizontal lines in the $2D$ plot. In the analysis that follows, we have 
only considered data measured sufficiently far from such modes. 

We observe that the visibility, defined as the minimum of each transmission 
dip, becomes enhanced on approaching the avoided level crossings. This phenomenon is visible in all experiments (see e.g. Fig. \ref{2Dplots_4K2}). 
It can be analyzed in more detail by looking at the frequency dependence of the transmission measured at fixed magnetic fields 
(Fig. \ref{2Dplot_650mK}b). Let's consider, for instance, the first 
transition, that links states with nuclear spin projection $m_{I} = 1/2$. 
The maximum absorption measured near the anticrossing, at $35.4$ mT and $9.3$ GHz, 
is approximately ten times larger than that measured away from it, at $53.7$ mT and $10$ GHz. The same comparison can be made, at constant $H$, 
between the intensities of different transitions that lie close or far from their respective avoided level crossings, e.g. the first and second 
transitions at $35.4$ mT (Fig. \ref{2Dplot_650mK}b).  

\begin{figure}
\centering
\includegraphics[width=\columnwidth]{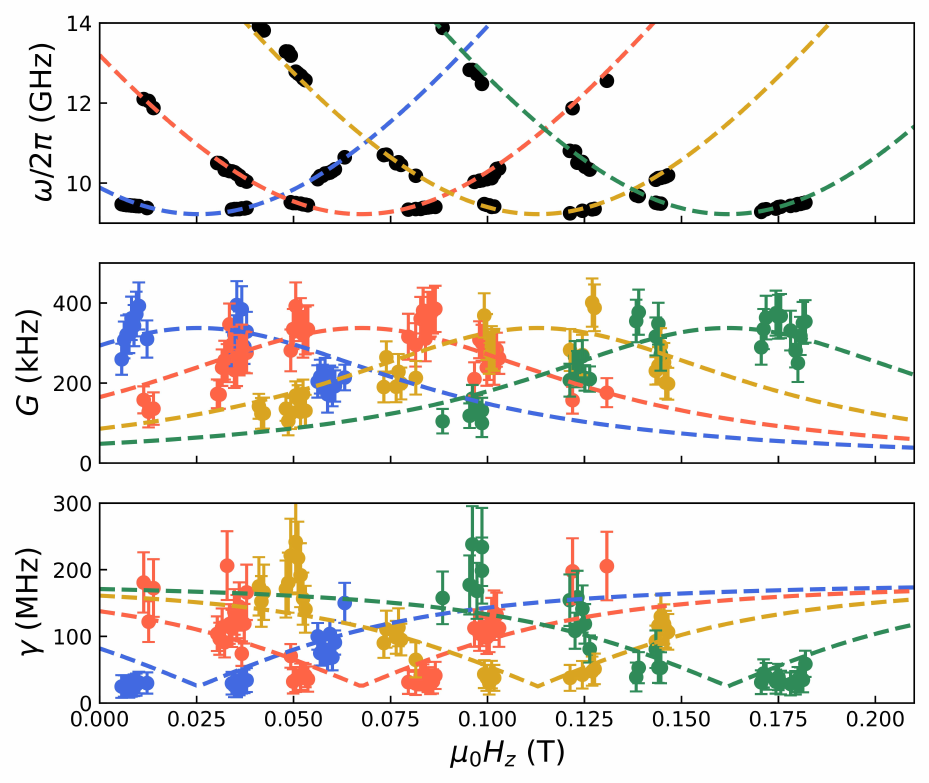}
\caption{Top: Frequencies of the spin resonances in Ho$_{0.2}$Y$_{0.8}$W$_{10}$ 
determined from microwave transmission experiments performed at $T=0.65$ K. The lines are $\omega_{1,2} = \sqrt{\Delta^{2} + [ 2g_{J} m_{J} (H_{z}-2m_{I}H_{z,1})]^{2}}$ with $g_{J}=5/4$, $m_{J}=4$, $\mu_{0} H_{z,1} = 23$ mT and $m_{I} = 1/2$ (blue), $3/2$ (red), $5/2$ (orange) and $7/2$ (green). Avoided 
level crossings lead to equally spaced minima as a function of magnetic field. 
Middle and bottom: Spin-photon coupling $G$ and resonance line 
width $\gamma$ obtained from the fit of these resonances by using Eq.(\ref{eq:transmission-line}). The lines in the middle pannel are calculated using Eq. (\ref{Gamma}) and the matrix element $\vert \langle \psi_{1)} \vert J_{z} \vert \psi_{2)} \rangle \vert^{2} = m_{J}^{2} \Delta^{2}/\omega_{12}^{2}$, which holds for a two-level 
system. The lines in the bottom panel are given by $\gamma = \gamma_{0} + b \partial \omega_{12}/ \partial H_{z}$ with $\gamma_{0} = 25$ MHz and $b = 4.5$ mT.}
\label{GvsH}
\end{figure}

In order to get a more quantitative characterization, fits of all absorption 
lines have been performed with Eq. (\ref{eq:transmission-line}). The 
fitting parameters were the spin-photon coupling $G$ and the line width 
$\gamma$, which here parameterizes the dominant inhomogeneous broadening. 
The results are shown in Fig. \ref{GvsH}. The increasing difficulty in properly normalizing the transmission plus the presence of a spurious mode near $9$ GHz prevents getting data right at the clock transitions. Yet, in 
spite of the experimental limitations, the results show that $G$ becomes maximum at the four avoided level crossings, as can be seen in Fig. \ref{GvsH}. Also, the linewidth seems to become larger on moving away from the anticrossings. 

This result admits a qualitative interpretation based on Eq. 
(\ref{Gamma}) and on the nature of the spin transitions in HoW$_{10}$ (see Fig. \ref{Levels}). The spin photon coupling is 
largely determined by the matrix element of $J_{z}$ between states 
with the same nuclear spin projections. The relevant subspace 
reduces then to a two-level tunneling system for which $\langle \psi_{1} \vert J_{z} \vert \psi_{2} \rangle \simeq \vert m_{J} \vert \Delta / \omega_{1,2}$, where $\omega_{1,2} = \sqrt{\Delta^{2} + [2g_{J} m_{J} (H_{z}-2m_{I}H_{z,1})]^{2}}$. The 
matrix 
element then inherits, although inverted, the field dependence of the level 
anticrossing. It reaches a maximum value $\langle \psi_{1} \vert J_{z} \vert \psi_{2} \rangle = \vert m_{J} \vert$ 
when the two levels come closest to each other ($\omega_{1,2} = \Delta$) and their 
wave functions become maximally delocalized between opposite angular momentum projections. Then, it decreases 
as the field moves away from $H_{n}$. Calculations performed inserting this simple expression for the matrix element into Eq. (\ref{Gamma}) reproduce quite well the experimental results, as shown in Fig. \ref{GvsH}. The 
maximum is enhanced in systems with a high 
spin ground state, as it is often the case with lanthanide ions and with HoW$_{10}$ in particular, for which $m_{J} = \pm 4$. 

Concerning the linewidth, it is expected that $\gamma$ decreases 
near the clock transitions. The electronic spins become then less sensitive 
to magnetic fields, thus also to perturbations arising from dipolar interactions with neighbour molecules \cite{Rubin-Osanz2021}, hyperfine 
couplings to nuclear spins \citep{Kundu2023} and the misalignment of the molecular axes. The 
experimental results, shown in the botom panel of Fig. \ref{GvsH}, confirm that spin resonances tend to narrow near the avoided level 
crossings. This effect can be approximately described 
by the expression $\gamma = \gamma_{0} + b \partial \omega_{12} /\partial H_{z}$,\cite{Balian2014,Shiddiq2016} where the second term is 
proportional to the effective magnetic moment, which tends to 
vanish near a clock transition and therefore suppress the effect of bias field broadening, and the first accounts for other sources of broadening. A reasonably good fit is obtained for $b \simeq 4.5$ mT  and $\gamma_{0} \leqslant 25-35$ MHz. The latter value turns out to be smaller than the level 
broadening, of about $120$ MHz, estimated from EPR experiments performed on diluted crystals and that was associated with a distribution in $B_{44}$.\cite{Shiddiq2016} The reason behind this discrepancy is not clear to us, 
but it might arise from a combination of the especial conditions of our  
experiments, which are sensitive to a tiny region within a tiny crystal, and the origin of the anisotropy parameter distribution. The results suggest that the line broadening we observe is dominated by environmental magnetic fields and that their influence is reduced near the level anticrossings. 

\section{Conclusions}

We have explored the coupling of HoW$_{10}$ molecular magnets 
to superconducting transmission waveguides. The results 
provide a broandband picture of the energy 
spectrum associated to the $m_{J} = \pm 4$ ground states. They confirm the existence of avoided level crossings, 
or spin clock transitions, at equispaced magnetic field values, 
determined by the magnetic anisotropy and hyperfine interactions. 
Near each anticrossing, we find that the spin-photon coupling 
$G$ becomes maximum, likely reflecting the maximum overlap between the two spin wavefunctions involved in the resonant transition. 
This reveals a quite unique property of spin-clock transitions. 
Not only do they shield spin states against magnetic field 
fluctuations, which leads to longer spin coherence times T$_{2}$,\cite{Shiddiq2016} 
but they also optimize their coupling to external radiation 
fields. Since $G \propto m_{J}^2$, the latter effect is enhanced 
in qubits that, like HoW$_{10}$, are characterized by a 
large effective ground state spin. This property makes spin clock 
transitions in artificial magnetic 
molecules highly promising for developing fast and robust spin 
qubits. The limitation imposed by the difficulty of tuning $\omega_{12}$ with a magnetic field can be compensated by exploiting electric fields, whose effect becomes maximum near the anticrossings.\cite{Liu2021}

The experimental scheme used in this work provides also the 
standard tool to control spin qubits on a chip, as has been shown 
by experiments performed on NV$^{-}$ centres in diamond \cite{Hanson2008} and impurity spins in silicon.\cite{Pla2012} Our  
results show a simple method to maximize the Rabi frequencies of single qubit operations on high-spin systems with suitable 
magnetic anisotropies. Besides, this scheme can be easily 
integrated with circuit QED architectures, e.g. with the 
application of superconducting resonators to read-out the spin 
states.\cite{Jenkins2016} The enhancement of the spin-photon 
coupling found here should also lead to larger dispersive shifts, 
thus improve the 
visibility of different spin states.\cite{Gomez-Leon2022} This is 
especially relevant when dealing with molecular spin qudits, e.g. 
those based on Gd$^{3+}$ ions.\cite{Jenkins2017} Even though Gd$^{3+}$ is a Kramers ion, the combination of non-diagonal magnetic anisotropy terms and adequately oriented external magnetic fields also leads to avoided level crossings in these systems. 
Exploiting the enhanced spin-photon coupling near them might then allow reaching the high cooperativity regime even with non too diluted crystals, and therefore provide a 
suitable platform for proof-of-concept implementations of qudit 
based algorithms.\cite{Chiesa2020,Chizzini2022}   

\section*{Acknowledgments}
This work has received support from grants RTI2018-096075-A-C21, PID2019-105552RB-C41, PID2019-105552RB-C44, P2018/NMT-4291 TEC2SPACE-CM, TED2021-131447B-C21, TED2021-131447B-C22, 
CEX2019-000919-M and CEX2020-001039-S, funded by MCIN/AEI/10.13039/501100011033, ERDF 'A way of making Europe' and ESF 'Investing in your future' and from the 
Gobierno de Arag\'on grant E09-17R-Q-MAD. We also acknowledge funding from the European Union Horizon 2020 research and innovation programme through FET-OPEN 
grant FATMOLS-No862893, 
ERC advanced grant Mol-2D-No788222, ERC consolidator grant DECRESIM-No647301 and 
HORIZON-MSCA-2021 grant HyQuArch-No101064707. This study forms also part of the 
Advanced Materials and Quantum Communication programmes, supported by MCIN with 
funding from European Union NextGenerationEU (PRTR-C17.I1), by Gobierno de Arag\'on, by Generalitat Valenciana and by CSIC (PTI001). SH acknowledges support of 
the US Department 
of Energy (DE-SC0020260). Work done at the National 
High Magnetic Field Laboratory is supported by the US 
National Science Foundation (DMR-1644779 and DMR-2128556) and the State of 
Florida.

\newpage

\bibliography{Gimeno_PhysRevApp_v2}

\end{document}